# Intrinsic nature of the giant spin Hall conductivity in Pt

Lijun Zhu[1*], Lujun Zhu[2], Manling Sui[3], Daniel C. Ralph[1,4], Robert A. Buhrman[1]

*1. Cornell University, Ithaca, NY 14850*

*2. College of Physics and Information Technology, Shaanxi Normal University, Xi'an, 710062, China*

*3. Institute of Microstructure and Property of Advanced Materials, Beijing University of Technology, Beijing 100124, China*

*4. Kavli Institute at Cornell, Ithaca, New York 14853, USA*

*lz442@cornell.edu

More than a decade after the first theoretical and experimental studies of the spin Hall conductivity (SHC) of Pt, both its amplitude and dominant origin remain in dispute. Resolving these questions is of fundamental importance for advancing understanding of very strong spin-orbit effects in conducting systems and for maximizing the spin Hall effect (SHE) for energy-efficient spintronics applications. Here we report the incorporation of MgO scattering centers to controllably vary the degree of long-range crystalline ordering of Pt films and thus its band structure. We establish that the dominant mechanism for the giant SHC in Pt is the intrinsic SHE of the Pt bulk, whereas its amplitude is considerably underestimated by the available theoretical calculations. Our work also establishes a compelling new spin Hall material $Pt_{0.6}(MgO)_{0.4}$ that combines a giant spin Hall ratio (> 0.3), with a low resistivity, a strong Dzyaloshinskii-Moriya interaction, and good integration compatibility for spintronics technology.

## INTRODUCTION

Since the first theoretical and experimental efforts on its spin Hall conductivity (SHC) a decade ago (*1-4*), platinum (Pt), the archetypal spin Hall material, has been central in generating and detecting pure spin currents and key in establishing most of the recent spin-orbit-coupling phenomena (*5-11*). However, the correct physical understanding of the SHC of Pt has remained unresolved, both qualitatively and quantitatively, despite the extensive attention that has been given to this intriguing condensed matter physics problem (*1-23*). The bulk spin Hall effect (SHE) of a heavy metal (HM) has three possible contributions, i.e. the intrinsic contribution from the band structure and the extrinsic skew-scattering and side-jump contributions from spin-orbit-interaction-related defect and impurity scattering. Theoretically, available first-principles calculations of the intrinsic SHC $\sigma_{SH}$ of Pt from the Berry curvature in the conduction band structure differ by more than a factor of 10 in the predicted room temperature values (i.e. $\sigma_{SH} = (0.4-4.5) \times 10^5$ $(\hbar/2e)\Omega^{-1} m^{-1})$(*1,13,14,18*), due to the markedly different predictions as to whether there is strong temperature dependence in $\sigma_{SH}$ (*2,18*). A tight-binding model calculation (*1*) for Pt predicts a 0-K Pt SHC that is a factor of ~1.7 less than that from first-principles calculations (*2*) and also predicts a degradation of the intrinsic SHC by decreasing quasi-particle lifetime (increasing impurity scattering), at first slowly, but then rapidly in the high resistivity range (e.g. the calculated intrinsic SHC decreases from $2.6\times10^5$, $1.6\times10^5$, to $0.1\times10^5$ $(\hbar/2e)$ $\Omega^{-1}$ m$^{-1}$ for Pt resistivity ($\rho_{xx}$) increases from 8, 65 to 200 $\mu\Omega$ cm)[1]. Experimentally, there is a strong disagreement as to both the strength (i.e. $(3.2\text{-}23.6)\times10^5$ $(\hbar/2e)\Omega^{-1}$m$^{-1}$)) and the dominant physical source for the SHC in Pt [15-21]. Some experiments are interpreted as indicating that the main contribution to the Pt SHC is extrinsic bulk skew scattering [19,20] and that therefore the spin Hall ratio ($\theta_{SH}=(2e/\hbar)\rho_{xx}\sigma_{SH}$) is independent of the resistivity ($\rho_{xx}$) with $\sigma_{SH}$ varying inversely with $\rho_{xx}$. Other recent works have examined the effect on $\theta_{SH}$ by alloying Pt [16,17] or by changing $\rho_{xx}$ of Pt via thickness [16,23] or temperature [15]. Both of these latter efforts report a close correlation between $\theta_{SH}$ and $\rho_{xx}$, consistent with either the side-jump or/and intrinsic contribution of the bulk SHE providing the dominant contribution to $\sigma_{SH}$ [15-17,23]. However, in the composition-dependent measurements of Pt alloys where a second heavy metal (HM) element is substituted into Pt lattice (e.g. $Au_{1-x}Pt_x$)[16], the definitive determination of the SHC mechanism is complicated by the fact that, in addition to $\rho_{xx}$ change (the disorder), there are simultaneous changes in the lattice constant (band structure), and the averaged bulk spin-orbit coupling strength. In the thickness and temperature dependent studies [15,16,23], the variation of $\sigma_{SH}$ with $\rho_{xx}$ cannot distinguish between intrinsic and side-jump contributions. Finally, there have also been experiments suggesting a significant interfacial SHC [21]. Clarifying the underlying physics of the giant SHC of Pt is both of fundamental interest and technological urgency (e.g. for maximizing $\theta_{SH}$ for low-power device applications [12-18]). Note that Pt and certain Pt-rich alloys [16,22] are, arguably, the most attractive class of spin Hall metals for energy-efficient applications because their combination of the highest SHC of any known class of metals with comparatively low $\rho_{xx}$ provides a giant $\theta_{SH}$ and minimal current shunting into an adjacent ferromagnet metal (FM) at the same time.

In this work, we report the results of a systematic variation of the degree of long-range crystalline ordering (LRCO) of a Pt film with increasing MgO impurity concentration, and therefore a tuning of its band structure perfection, and a concomitant variation of its $\rho_{xx}$. We find that the SHC varies directly, albeit at first only weakly, with the degree of LRCO of the Pt lattice. We also find that dampinglike spin-orbit torque (SOT) and thus $\theta_{SH}$ scale closely with $\rho_{xx}$, but not exactly due to the gradual changes in the SHC that occur along with the strong increase in resistivity as the LRCO is reduced. We argue that these results conclusively establish that the dominant source of the giant SHC in Pt is the intrinsic contribution related to the Berry curvature in the Pt band structure. We also emphasize that the large experimental SHC values for Pt and its alloys indicate that the existing first-principles and tight-binding theories [1,2,12-14,18] are underestimating the true intrinsic SHC in Pt, particularly so if model calculations of spin backflow and/or spin memory loss at the Pt/Co interface are at least approximately applicable to the experimental situation. Finally, by utilizing this intrinsic SHE mechanism of Pt, we achieve a 100% enhancement in the spin-orbit torque (SOT), thus $\theta_{SH}$, by increasing $\rho_{xx}$ via the incorporation of MgO scatterers into Pt. This establishes a new ternary spin Hall material $Pt_{0.6}(MgO)_{0.4}$ that is very compelling for low-power SOT device technology because of its combination of a giant $\theta_{SH}$, a low $\rho_{xx}$, a large Dzyaloshinskii-Moriya interaction (DMI), and easy growth on silicon substrates by sputtering.

## RESULTS AND DISCUSSION

### Sample structure and resistivity

Multilayer stacks of Ta 1.0/$Pt_{1-x}(MgO)_x$ 4.0 /Co 0.68-1.4/MgO 2.0/Ta 1.5 (numbers are thickness in nm) with $x =$ 0, 0.25, 0.5, 0.65, 0.75, 0.85, and 1, respectively, were sputter deposited at room temperature on Si/SiO₂ substrates. The $Pt_{1-x}(MgO)_x$ layer was cosputtered from a Pt target and a MgO target. The Co magnetization ($M_s$) for these samples was measured by a vibrating sample magnetometer (VSM) to be ~1220±85 emu/cc, indicating the absence of a significant magnetic proximity effect (22) in these as-grown samples. High-resolution cross-sectional scanning transmission electron microscopy (STEM) studies of the Pt(MgO) composites (see Fig. 1(b) for images of a $Pt_{0.6}(MgO)_{0.4}$ sample) show that the co-sputtered $Pt_{1-x}(MgO)_x$ layer has a homogeneous polycrystalline texture where the grains are ~ 4 nm in vertical extent and that there is no indication of the presence of MgO clusters of observable size. Figure 1(c) shows cross-sectional energy-dispersive x-ray spectroscopy (EDS) Mg and O mapping of the composite material (Fig. 1(c)) under the HAADF mode, which supports, within the resolution, the conclusion that the MgO is finely dispersed within the Pt. We find from x-ray photoemission spectroscopy (XPS) studies (Fig. 2(a)) that the Pt $4f_{7/2}$ and $4f_{5/2}$ peaks are located at 71.1 and 74.4 eV, respectively, in both a $Pt_{0.6}(MgO)_{0.4}$ and a pure Pt layer. In contrast, the binding energies of $4f_{7/2}$ and $4f_{5/2}$ peaks for Pt oxides are reported to be shifted to ~72.3 and ~75.8 eV for $Pt^{2+}$, and 74.0 and 77.5 eV for $Pt^{4+}$, respectively (24,25). Meanwhile, we find that the XPS peaks for Mg KLL and O 1s are shifted by ≈ ± 0.8 eV, respectively (fig. S1). This clearly indicates that the Pt atoms in the $Pt_{1-x}(MgO)_x$ layers are not oxidized while the Mg atoms are oxidized, consistent with the fact that Pt has a much stronger electronegativity than Mg.

Figure 2(b) shows x-ray diffraction (XRD) $\theta$-$2\theta$ patterns of $Pt_{1-x}(MgO)_x$ 4/Co 1.4 bilayers with different MgO concentration $x$. Similar to the case of $Au_{1-x}Pt_x$ alloys(16,22), the $Pt_{1-x}(MgO)_x$ layer shows a broad face-centered-cubic (fcc) (111) peak due to the polycrystalline texture and the small thickness. However, unlike the case for Pt alloyed with another fcc metal, the $Pt_{1-x}(MgO)_x$ (111) peak does not shift with $x$, despite the dramatic broadening with increasing MgO concentration. This indicates that the MgO molecules are primarily dispersed in the Pt as inter-site impurities rather than being substituted into the Pt lattice (16,22).

As shown in Figs. 2(c)-2(e), the decreasing integrated intensity and the increasing full width at half maximum (FWHM) of the $Pt_{1-x}(MgO)_x$ (111) diffraction peak clearly indicate that the LRCO of the $Pt_{1-x}(MgO)_x$ layers is monotonically weakened by MgO incorporation. We note here that in this XRD measurement configuration where the grain size along the film normal is the film thickness (~4nm, see Fig. 1(b)) the variation of the FWHM is dominated by a variation in LRCO that is not due to a variation of the Pt grain size. While the LRCO of the fcc crystalline lattice of Pt is clearly gradually diminished by the increasing MgO content, it is also somewhat surprising how resilient the lattice periodicity is against a larger degree of MgO incorporation, until a sharp structural degradation occur at $x = 0.5$. This certainly indicates that the Pt band structure, which is determined by the LRCO, is only disrupted slowly toward that of amorphous Pt with increasing addition of atomic scale interstitial defects. The average resistivity for $Pt_{1-x}(MgO)_x$ was determined for each $x$ by measuring the conductance enhancement of the corresponding stacks with respect to a reference stack with no $Pt_{1-x}(MgO)_x$ layer. As summarized in Fig. 2(e), $\rho_{xx}$ for the Pt$_{1-x}$(MgO)$_x$ increases gradually from 33 μΩ cm at $x = 0$ (pure Pt) to 74 μΩ cm for $x = 0.4$, and then jumps up to 240 μΩ cm, which is qualitatively consistent with the variation in the degree of LRCO indicated by the intensity and FWHM of XRD (111) peaks.

**Determination of effective spin-orbit torque fields**

A lock-in amplifier was used to source a sinusoidal voltage ($V_{in} = 4$ V) across the length of the bar ($L = 60$ μm) orientated along the $x$ axis (see Fig. 3(a)) and to detect the in-phase first and out-of-phase second harmonic Hall voltages, $V_{1\omega}$ and $V_{2\omega}$. For in-plane magnetized HM/FM bilayer ([16,26]), the dependence of $V_{2\omega}$ on the in-plane field angle ($\varphi$) is given by $V_{2\omega} = V_a\cos\varphi + V_p\cos\varphi\cos2\varphi$, where $V_a = -V_{AH}H_{DL}/2(H_{in}+H_k)+V_{ANE}$, and $V_p = -V_{PH}H_{FL}/2H_{in}$ with $V_{AH}$, $V_{ANE}$, $H_{in}$, $H_k$, $V_{PH}$, $H_{DL}$, and $H_{FL}$ being anomalous Hall voltage, anomalous Nerst voltage ([26]), in-plane bias field, perpendicualr anisotropy field, planar Hall voltage, damping-like effective spin torque field, and field-like effective spin torque field. The $\varphi$ dependence of $V_{1\omega}$ is given by $V_{1\omega} = V_{PH}\cos2\varphi$, from which the planar Hall voltage $V_{PH}$ can be determined (see Fig. 3(b)). We separated the damping-like term $V_a$ and the field-like term $V_p$ for each $H_{in}$ and each $x$ by fitting $V_{2\omega}$ data to $V_{2\omega} = V_a\cos\varphi + V_p\cos\varphi\cos2\varphi$ (see Fig. 3(c)). The linear fits of $V_a$ versus $-V_{AH}/2(H_{in}+H_k)$ and $V_p$ versus $-V_{PH}/2H_{in}$ (see Fig. 3(d) and 3(e)) yield the values of $H_{DL}$ and $H_{FL}$ for Pt$_{1-x}$(MgO)$_x$/Co bilayers. Figure 4(a) summarizes $H_{DL}$ and $H_{FL}$ for a constant applied electric field $E$ $=L/V_{in}$=66.7 kV m$^{-1}$. Both $H_{DL}$ and $H_{FL}$ vary significantly and monotonically with $x$.

**Physical origin of the giant SHC in Pt**

Using the experimental results for the dampinglike effective SOT fields (Fig. 4(a)), we calculated the *apparent* or effective spin Hall conductivity $\sigma_{SH}^* \equiv T_{int}\sigma_{SH} = (\hbar/2e)\mu_0 M_s t H_{DL}/E$ as a function of $x$ in Fig. 4(b). Here $T_{int}$ ($< 1$) is the spin transparency of the HM/FM interface is set by spin backflow ([27]) and spin memory loss scattering at the interfac ([28,29]); $e$, $\mu_0$, $t$, and $\hbar$ are the elementary charge, the permeability of vacuum, the ferromagnetic layer thickness, and the reduced Planck constant, respectively. For the purposes of comparison we also plot the Pt$_{1-x}$(MgO)$_x$ conductivity $\sigma_{xx}(x)$ in Fig. 4(b). With the addition of increasing amounts of MgO, $\sigma_{SH}^*(x)$ decreases only gradually from 4.9 ×10$^5$ ($\hbar$/2e) Ω$^{-1}$ m$^{-1}$ at $x$ =0 to 4.1 ×10$^5$ ($\hbar$/2e) Ω$^{-1}$ m$^{-1}$ at $x$ = 0.4 (a 16 % decrease), and then sharply drops to 1.3×10$^5$ ($\hbar$/2e) Ω$^{-1}$ m$^{-1}$ at $x$ = 0.5. We note that for $x \leq 0.4$ the percentage decrease in $\sigma_{xx}(x)$ is much greater (by ~56%) than that of $\sigma_{SH}^*(x)$(by ~16%). This obviously disagrees with the skew scattering mechanism ($\sigma_{SH} \propto \sigma_{xx}$) but is consistent with a dominant intrinsic SHC mechanism where $\sigma_{SH}^*(x)$ arises from the LRCO modified by the MgO addition instead of the density of elastic scattering centers ($\sigma_{xx}$). Our result is also qualitatively consistent with the tight-binding model prediction ([1]) that the Pt band structure components that give rise to the Berry curvature are relatively robust in the presence of moderate disorder and thus that $\sigma_{SH}^*$ should initially only vary slowly with decreasing $\sigma_{xx}$, before decreasing much more rapidly once $\sigma_{xx}$ is below a critical threshold. Quantitatively, however, it appears that the Pt Berry curvature is considerably more resilient than predicted against increases in the scattering rate caused by the increased structural disorder over the entire range of resistivity (MgO concentration) that we studied (33 μΩ cm to 240 μΩ cm.) We also point out that the clear qualitative correlation between the evolution of $\sigma_{SH}^*(x)$ over the entire range ($0 < x \leq 0.5$) and the changes in the LRCO of Pt bulk indicated by the FWHM and XRD intensity (Figs. 2(c) and 2(d)) confirms that the spin torques that we observe in this system are the result of a bulk effect (i.e. due to the SHC of the $Pt_{1-x}(MgO)_x$ layer) rather an interfacial effect. Thus previous reports of exceptionally strong SOTs from the interfaces of the oxides ($WO_x$ (30), $CuO_x$ (31), and $PtO_x$ (24)) are not relevant to our $Pt_{1-x}(MgO)_x$ /Co case. This conclusion of a predominant bulk origin of the SOTs in our Pt-based system is consistent with the previously reported rapid decrease of the SOTs strength ($\sigma_{SH}^*(x)$) with the HM thickness when the HM thickness becomes less than the effective spin-diffusion length (16,23).

Typically, the dampinglike and fieldlike SOT efficiencies per unit bias current density, $\xi_{DL}$ and $\xi_{FL}$, or the (effective) spin Hall ratio (angle) are reported in SOT experiments because the efficiency, by which a bias current in the HM generates a spin current that exerts torques on an adjacent FM layer, is the most direct parameter that characterizes the useful strength of the phenomenon. In Figure 4(c) we show the results for $\xi_{DL}$ and $\xi_{FL}$ as calculated from the effective field measurements by $\xi_{DL(FL)} = 2e\mu_0 M_s t H_{DL(FL)}/\hbar j_e$ (=$(2e/\hbar)$ $\sigma_{SH}^* \rho_{xx}$), with $j_e = E/\rho_{xx}$ being the charge current density in the $Pt_{1-x}(MgO)_x$ layer. Since $\sigma_{SH}^*$ varies only moderately for $x \leq 0.4$, $\xi_{DL}$ scales rather closely with $\rho_{xx}$, increasing from 0.16 at $x = 0$ to 0.30 at $x = 0.4$. At $x = 0.5$, despite that $\rho_{xx}$ increases sharply to 240 μΩ cm, $\xi_{DL}$ only slightly increases to 0.31 due to strong decrease in $\sigma_{SH}$ as the result of the sharp increase in structural disorder. This large change in $\sigma_{SH}$ breaks the approximate scaling between $\xi_{DL}$ and $\rho_{xx}$ that is followed for $x \leq 0.4$. We note that $\xi_{DL}$ for the 4 nm pure Pt sample here (0.16) is comparable to that reported previously by our group for Pt with a similar thickness (0.12-0.18)(16,23,28). The slight variation in $\xi_{DL}$ is mainly attributed to the differences in resistivity and the interfacial spin memory loss resulting from film growth protocols. We also note that, qualitatively, the quasi-linear scaling of $\xi_{DL}$ with $\rho_{xx}$ is in accord with that found previously with light Pt alloying (16,17). These results clearly reaffirms the absence of any important skew scattering to the bulk SHC of Pt and non-magnetic Pt alloys, because in that case one would expect $\xi_{DL}$ to be approximately independent of $\rho_{xx}$.

With regard to the possibility of an extrinsic side-jump contribution to the $\sigma_{SH}$ of Pt and $Pt_{1-x}(MgO)_x$, we can draw from the conclusions of previous studies of side-jump contribution to the anomalous Hall conductivity (32). Such research indicates that the side jump contribution to $\sigma_{SH}$ should be expected to scale inversely with the square of the residual resistivity ratio (RRR = $\rho_{xx}/\rho_{xx0}$) of the metal, where $\rho_{xx0}$ is the residual, low temperature resistivity due to elastic, defect scattering. As shown in fig. S2, the RRR for our pure Pt sample ($x = 0$) is approximately 1.4, while for $x = 0.4$ it is $\approx 1.1$. In light of the small *decrease* in $\sigma_{SH}$ between $x = 0$ and $x = 0.4$, this is certainly not consistent with side jump being a significant contributor to the SHC of Pt and $Pt_{1-x}(MgO)_x$. Therefore, we conclude that the intrinsic contribution is dominant physical origin of the observed SHC in Pt and SOTs in Pt/FM systems.

It is important to note that the experimental values we obtain for $\sigma_{SH}^*$ are equal or larger than the available theoretical predictions for Pt from first-principles or tight-binding calculations. Moreover, these experimental values of the *apparent* SHC $\sigma_{SH}^*$ measured in our $Pt_{1-x}(MgO)_x$ samples with $x \leq 0.4$ (e.g. $4.9 \times 10^5$ $(\hbar/2e)\Omega^{-1}$ m$^{-1}$ for $x = 0$) do not take into account the unavoidable spin current loss at the interface due to the spin backflow and the possible interfacial spin memory loss. If we only consider the ideal situation of spin backflow being important, the available band structure calculations for Pt/Co predict an interfacial spin transparency $T_{int} \approx 0.5$ (*19,23,28*). Therefore, if the backflow analysis is applicable, the actual intrinsic SHC of Pt is then at least $\sim 1.0 \times 10^6$ $(\hbar/2e)$ $\Omega^{-1}$ m$^{-1}$. This is significantly larger than any of the existing theoretical predictions from first-principles or tight-binding calculations: $(0.4\text{-}4.5) \times 10^5$ $(\hbar/2e)$ $\Omega^{-1}$ m$^{-1}$ (*1,2, 12-14,18*), indicating that the available calculations are underestimating the true intrinsic SHC of Pt by at least a factor of 2, and most likely more since there is significant evidence of substantial spin memory loss at Pt/Co interfaces (29). We infer that there is still important underlying physics related to the generation of spin currents by intrinsic effects in Pt that is yet to be fully understood and that could benefit from additional theoretical investigation.

**Practical impact for low-power SOT devices**

The giant $\xi_{DL}$ of 0.30 for the 4 nm $Pt_{0.6}(MgO)_{0.4}$ ($\rho_{xx} = 74$ μΩ cm) is comparable to the high value reported for fcc-$Au_{0.25}Pt_{0.75}$ ($\rho_{xx} \sim 83$ μΩ cm)(*16*) and $\beta$-W ($\rho_{xx} \sim 300$ μΩ cm)(*33*) and 3 times higher than that of $\beta$-Ta ($\rho_{xx} \sim 190$ μΩ cm)(*34*).The SHE in those HMs has been demonstrated to enable sub-ns deterministic magnetic memories (*8,35*), gigahertz and terahertz oscillators (*9,10*), and fast skyrmion/chiral domain wall devices (*11,36*). However, for low-power device applications, new HMs that simultaneously combine a giant $\xi_{DL}$ with a low $\rho_{xx}$ and a good compatibility for device integration are still urgently required (*16*). In that regard, we first point out that $Pt_{0.6}(MgO)_{0.4}$ ($\xi_{DL} = 0.30$ and $= \rho_{xx} = 74$ μΩ cm) is somewhat more energy-efficient than $Au_{0.25}Pt_{0.75}$(*16*), and progressively more so than Pt, $\beta$-W (*33*), $\beta$-Ta (*34*) and the topological insulator $Bi_xSe_{1-x}$ ($\xi_{DL}=3.5\text{-}18.6$) (*37,38*) for SOT applications with metallic magnets, e.g. in-plane magnetized FeCoB-MgO MRAMs (see quantitative comparison in Table 1), after taking into account the current shunting into the ferromagnetic layer (see section S1 for details on the power calculations). The relatively small $\rho_{xx}$ of the $Pt_{0.6}(MgO)_{0.4}$ is also highly desirable for applications that require high energy efficiency but small write impedance, e.g. the prospective implementation of SOT devices in cryogenic computing systems (*39*). In that case, the very resistive $\beta$-W (*33*), $\beta$-Ta (*34*), and $Bi_xSe_{1-x}$(*37,38*) spin Hall materials are all problematic.

As an independent check of the validity of the strong damping-like SOT generated by the $Pt_{1-x}$ (MgO)$_x$ as measured by the in-plane harmonic response technique, we show in Figs. 5(a) and 5(b) the deterministic switching of the magnetization of a 0.68 nm thick perpendicularly magnetized Co layer by the strong damping-like SOT generated by the SHE in a 4 nm $Pt_{0.7}(MgO)_{0.3}$ layer (we use $Pt_{0.7}(MgO)_{0.3}$ rather than $Pt_{0.6}(MgO)_{0.4}$ to provide stronger perpendicular magnetic anisotropy and larger coercivity). In this measurement, the dc current was sourced by a Yokogawa 7651 and the differential Hall resistance was detected by the lock-in amplifier ($V_{in} = 0.1$ V). The Co layer has a coercivity ($H_c$) of ~80 Oe (Fig. 5(a)) and effective perpendicular anisotropy field ($H_k$) of ~2.0 T as determined by fitting the dependence of $V_{1\omega}$ on the in-plane bias field $H_x$ following the parabolic relation $V_{1\omega} = \pm V_{AH}(1 - H_x^2/2H_k^2)$ (see Fig. 5(b)). As shown in Fig. 5(c) and 5(d), the dampinglike SOT generated by the SHE of $Pt_{0.7}(MgO)_{0.3}$ layer enables the sharp deterministic switching of a perpendicularly magnetized Co layer through domain wall depinning at a switching current of ~2.7 mA, which corresponds to a critical switching current density of $j_e = 1.15 \times 10^7$ A/cm$^2$ in the $Pt_{0.7}(MgO)_{0.3}$ layer. To obtain reliable switching in this configuration it was necessary to apply an in-plane magnetic field of > 1000 Oe along the current direct to overcome the domain wall chirality imposed by the Dzyaloshinshii-Moriya interaction (DMI) at the HM/Co interface (*40*).

Here we also note that the requirement of a large in-plane bias field (>1000 Oe) for switching perpendicular magnetization indicates a strong DMI at the $Pt_{1-x}(MgO)_x$/FM interface, which, together with the high $\xi_{DL}$ (Fig. 4), makes $Pt_{1-x}(MgO)_x$ particularly attractive for energy-efficient skyrmion and chiral domain wall devices driven by SHE-governed domain wall depinning (*11*). However, such a bias field is not required for anti-damping SOT switching of collinear or "*y*-type" in-plane magnetized magnetic memories (*41*), which have been demonstrated to be surprisingly efficient and fast (e.g. critical switching current (density) ~110 μA (~5×10$^6$ A/cm$^2$), ~200 ps for W-based FeCoB-MgO MRAM devices (*35*)).

**Discussion**

In conclusion, we have clarified that the origin of the giant SHC in Pt is the intrinsic SHE arising from the band structure of the Pt bulk by systematically tuning the perfection of the crystalline and thus the electronic structure via incorporation of MgO inter-site scattering centers into the Pt bulk. The conclusion that the SHE in Pt and $Pt_{1-x}(MgO)_x$ is dominated by a very robust intrinsic band structure is clearly validated by the qualitative correlation between $\sigma_{SH}$ and the degree of LRCO and the quantitative scaling of $\xi_{DL}$ and $\rho_{xx}$, which together exclude any important skew-scattering, side-jump, or interfacial contributions to the spin Hall effect in this system. Moreover, the internal SHC is experimentally found to be > $1 \times 10^6$ ($\hbar/2e$) $\Omega^{-1}$ m$^{-1}$, a value that is considerably underestimated by the existing first-principle and tight-binding theories. By utilizing the intrinsic nature of the SHE in Pt, we obtained a 100% enhancement of the dampinglike SOT efficiency $\xi_{DL}$ by increasing Pt resistivity via incorporating MgO impurities. This establishes a new spin Hall material $Pt_{0.6}(MgO)_{0.4}$ that is very compelling for low-power SOT applications in magnetic memories, oscillators, and skyrmion/chiral domain wall devices due to its combination of a giant spin Hall angle ($\theta_{SH} > \xi_{DL} = 0.3$) with a relatively low resistivity (~74 μΩ cm), a strong DMI, and a good compatibility with the fabrication requirements for integrated circuit technologies.

**MATERIALS AND METHODS**

All the samples were sputter deposited at room temperature on Si/SiO$_2$ substrates with argon pressure of 2 mTorr and a base pressure of below $1 \times 10^{-8}$ Torr. The $Pt_{1-x}(MgO)_x$ layers are co-sputtered from a Pt target and a MgO target. The 1 nm Ta underlayer was introduced to improve the adhesion and the uniformity of the stack. The 1.5 nm Ta capping layer was fully oxidized upon exposure to atmosphere. The sample structure was first characterized by combining cross-sectional high-resolution scanning transmission electron microscopy (STEM) imaging, high-angle annular dark field (HAADF) STEM imaging, and energy-dispersive x-ray spectroscopy (EDS) mapping in a spherical-aberration-corrected (Cs-corrected) 300-kV FEI Titan G2 microscope equipped with a Super-X detector. Focused ion beam (FIB, FEI Helios Nanolab 600i) was used during the preparation of the STEM samples. The stacks were patterned into 5×60 $\mu m^2$ Hall bars by ultraviolet photolithography and argon ion milling for harmonic response measurements and direct current switching experiments. The magnetization of Co layers and the chemical bond information in the $Pt_{1-x}(MgO)_x$ layers are measured by vibrating sample magnetometer and x-ray photoemission spectroscopy, respectively.

## SUPPLEMENTARY MATERIALS

Supplementary material for this article is available at XXXXXXXX

fig. S1. XPS spectrum for Pt $4d_{5/2}$, Pt $4d_{3/2}$, Pt $4p_{3/2}$, Mg KLL, and O 1s in a $Pt_{0.6}(MgO)_{0.4}$ layer.

fig. S2. Temperature dependence of resistivity for 4-nm-thick Pt and $Pt_{0.6}(MgO)_{0.4}$ films.

fig. S3. Schematics for a spin-orbit torque MRAM device.

section S1. Calculation of power consumption of spin-orbit torque MRAM devices.

**Acknowledgments**: This work was supported in part by the Office of Naval Research (N00014-15-1-2449), by the NSF MRSEC program (DMR-1719875) through the Cornell Center for Materials Research, and by the Office of the Director of National Intelligence (ODNI), Intelligence Advanced Research Projects Activity (IARPA), via contract W911NF-14-C0089. The views and conclusions contained herein are those of the authors and should not be interpreted as necessarily representing the official policies or endorsements, either expressed or implied, of the ODNI, IARPA, or the U.S. Government. The U.S. Government is authorized to reproduce and distribute reprints for Governmental purposes notwithstanding any copyright annotation thereon. This work was performed in part at the Cornell NanoScale Facility, an NNCI member supported by NSF Grant ECCS-1542081.

**Author contributions**: Lijun Zhu and R. A. Buhrman designed the project and wrote the draft. Lijun Zhu fabricated the samples, performed sample fabrication, VSM, XRD, XPS, and spin-torque measurements. Lujun Zhu and Manling Sui performed the TEM measurements. All the authors contributed to discussing the results and writing the manuscript. **Competing interests**: The authors declare that they have no competing interests.

**Data and materials availability**: All data needed to evaluate the conclusions in the paper are present in the paper and/or the Supplementary. Additional data related to this paper may be requested from the authors.

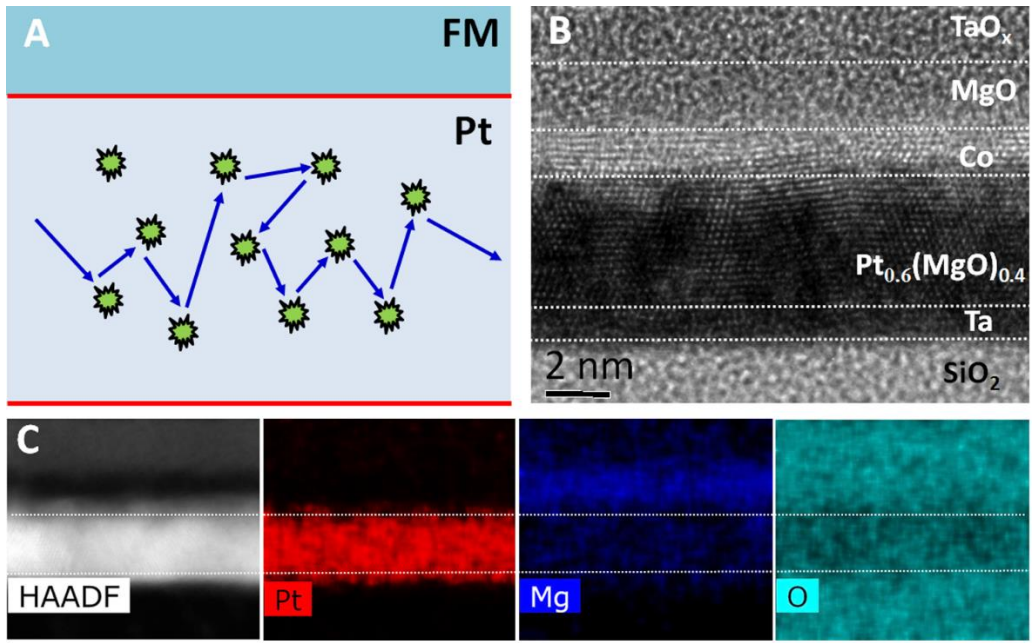

**Fig. 1. Cross-sectional sample structure**. (A) Schematic of enhanced electron scattering in Pt by MgO impurities; (B) Cross-sectional high-resolution STEM image (bright field) of a magnetic stack of Ta 1/ $Pt_{0.6}(MgO)_{0.4}$ 4/Co 1.4/MgO 2/TaO$_x$ 1.5; (C) Cross-sectional HAADF image and EDS mapping of Pt, Mg, and O. The two dashed lines represent the upper and lower interfaces of the $Pt_{1-x}(MgO)_x$ layer.

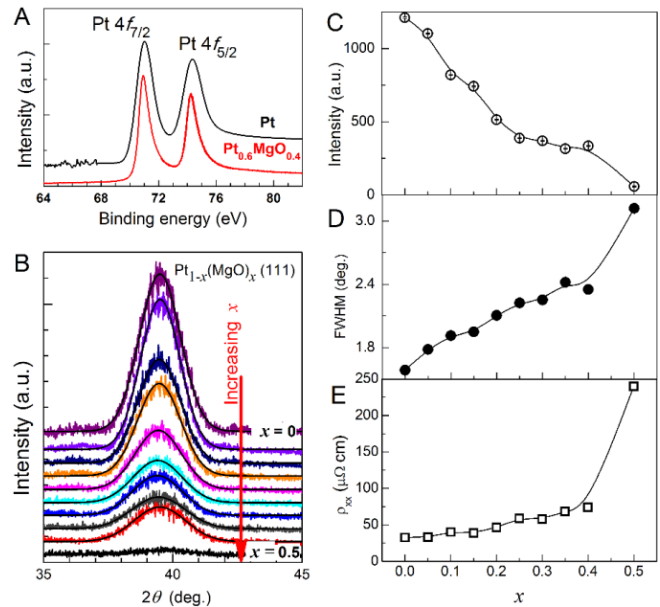

**Fig. 2. Degradation of LRCO and enhancement of scattering of Pt by MgO incorporation**. (A) XPS spectrum for Pt $4f$ peaks in a $Pt_{0.6}(MgO)_{0.4}$ layer (black line) and a pure Pt layer (red line), indicating non-oxidization of Pt in both cases; (B) XRD $\theta$-$2\theta$ patterns for $Pt_{1-x}(MgO)_x$ with different $x$. (C)-(E) integrated intensity of the $Pt_{1-x}(MgO)_x$ (111) peak, FWHM of the $Pt_{1-x}(MgO)_x$ (111) peak, and the average resistivity plotted as a function of Pt concentration $x$.

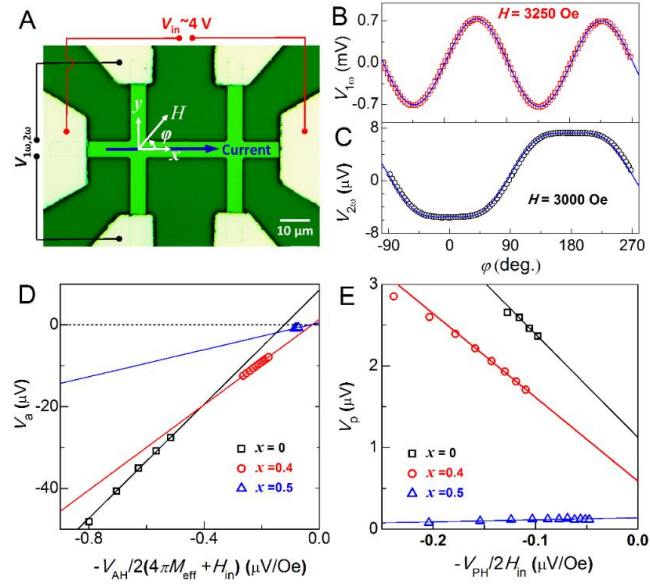

**Fig. 3. Determination of Spin-orbit torques.** (A) Scanning electron microscopy image of Hall bar devices; (B) $\varphi$ dependence of $V_{1\omega}$ ($x = 0.6$), (b) $\varphi$ dependence of $V_{2\omega}$ ($x = 0.6$), (C) $V_a$ versus $-V_{AH}/2(H_{in}+H_k)$, and (d) $V_p$ versus $-V_{PH}/2H_{in}$ for $Pt_{1-x}(MgO)_x$/Co bilayers with different $x$. The solid lines in (B)-(D) represent best fits of data to $V_{1\omega} = V_{PH}\cos2\varphi$, $V_{2\omega} = V_a\cos\varphi + V_p\cos\varphi\cos2\varphi$, $V_a = -V_{AH}H_{DL}/2(H_{in}+H_k)+V_{ANE}$, and $V_p = -V_{PH}H_{FL}/2H_{in}+$offset, respectively.

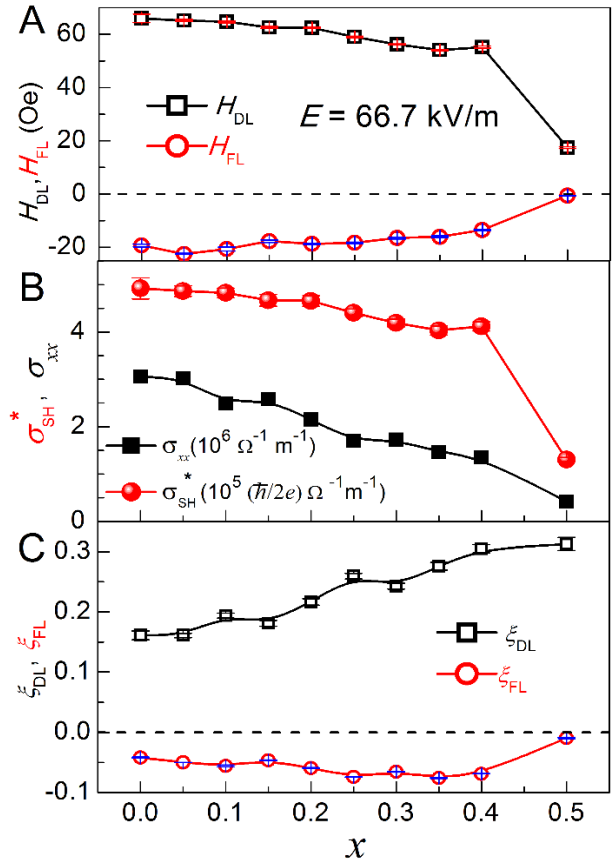

**Fig. 4. MgO concentration dependence of spin-orbit torques**. (A) $H_{DL(FL)}$, (B) $\sigma_{SH}^*$, (C) $\xi_{DL(FL)}$.

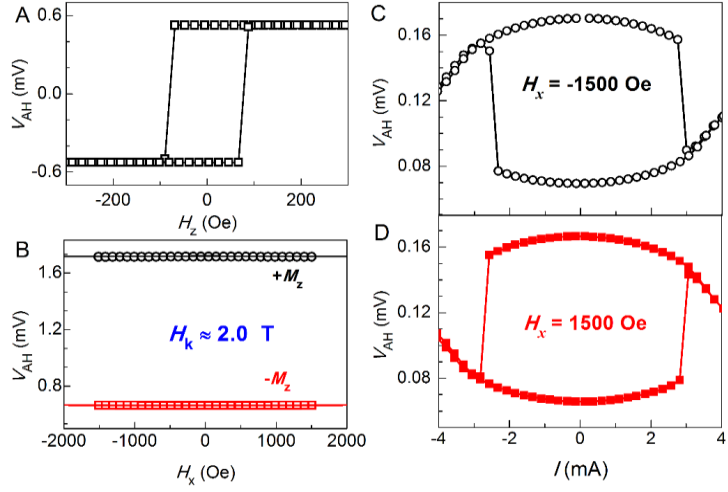

**Fig. 5. Deterministic spin-orbit torque switching of a perpendicular magnetization in a Pt$_{0.7}$(MgO)$_{0.3}$ 4/Co 0.68 bilayer**. (A) $V_{AH}$ versus $H_z$, (B) $V_{AH}$ versus $H_x$, (C) $V_{AH}$ versus $I$ ($H_x$ = -1500 Oe), (D) $V_{AH}$ versus $I$ ($H_x$ = 1500 Oe), respectively. The Hall bar dimension is 5 ×60 μm$^2$.

**Table I. Comparison of $\xi_{DL}$, $\rho_{xx}$, $\sigma_{SH}^*$, and normalized power consumption of SOT-MRAM devices for various strong spin current generators**. Here we use a 600 ×300 ×4 nm$^2$ spin Hall channel, a 190 ×30 ×1.8 nm$^3$ FeCoB free layer (resistivity ≈130 μΩ cm) and the parallel resistor model for the illustrative calculation (see Supporting information for details on the power calculations).

| | $\xi_{DL}$ | $\rho_{xx}$ (μΩ cm) | $\sigma_{SH}^*$ ($10^5$ $(\hbar/2e)\Omega^{-1}m^{-1}$) | Power | Ref. |
|---|---|---|---|---|---|
| Pt$_{0.6}$(MgO)$_{0.4}$ | 0.3 | 74 | 4.1 | 1.0 | This work |
| Au$_{0.25}$Pt$_{0.75}$ | 0.3 | 83 | 3.6 | 1.2 | Zhu *et al* [16] |
| Pt | 0.16 | 33 | 4.9 | 1.3 | This work |
| Bi$_2$Se$_3$ | 3.5 | 1755 | 2.0 | 2.9 | Mellnik *et al* [37] |
| $\beta$-W | 0.3 | 300 | 1.0 | 8.1 | Pai *et al* [33] |
| $\beta$-Ta | 0.12 | 190 | 0.63 | 24 | Liu *et al* [34] |
| Bi$_x$Se$_{1-x}$ | 18.6 | 13000 | 1.4 | 28 | DC *et al* [38] |